\documentstyle[12pt]{article}
\begin{document}

\baselineskip24pt
\vbox to 2cm{}

\begin{center}
{\large{\bf Phase-Space Exploration in \\
Nuclear Giant Resonance Decay}}
\vspace{1.0cm}

S. Dro\.zd\.z$^{a,b,c}$,
S. Nishizaki$^{a,d}$,
J. Wambach$^{a,c}$
and J. Speth$^{a}$
\vspace{1.0cm}

{\it
 a) Institut f\"ur Kernphysik, Forschungszentrum J\"ulich,
D-52425 J\"ulich, Germany \newline
 b) Institute of Nuclear Physics, PL - 31-342 Krak\'ow, Poland \newline
 c) Department of Physics, University of Illinois at Urbana,
IL 61801, USA\newline
 d) College of Humanities
and Social Sciences, Iwate University, Ueda  3-18-34,\newline
Morioka 020, Japan}

\vskip 1cm

{\large Abstract}

\end{center}

The rate of phase-space exploration in the decay of isovector
and isoscalar giant quadrupole resonances in $^{40}$Ca is analyzed.
The study is based on the time dependence of the survival probability
and of the spectrum of generalized entropies evaluated in the space
of 1p-1h and 2p-2h states.
If the 2p-2h background shows the characteristics
typical for chaotic systems, the isovector excitation evolves almost
statistically while the isoscalar excitation remains largely localized,
even though it penetrates the whole available phase space.

\vspace{1.0cm}

\vfill

\newpage

Collective nuclear excitations occur on dynamical time scales which are
short compared to those of the compound nucleus and,
therefore, probe simple configurations. In response to an external
perturbation a collective mode is initially formed as
a non-stationary state which occupies a small fraction of the available
phase space. The subsequent decay, on the other hand, involves much
longer time scales and explores more complex nuclear configurations.
The specific characteristics of this process may, of course, depend
on the energy distribution of the initial state, its multipolarity,
its isospin, character or the degree of collectivity.
The generic features of the time-dependent phase-space
exploration through decay, however, are of much more
general nature \cite{Hel} and relate to the quantum manifestation
of classical chaos in the time-dependent picture \cite{LLJP,WB,LQSL,AW}.
The nucleus is especially well suited for addressing such questions
because of the
inherent quantum nature and a generic chaoticity of the dynamics.
There is also
a wealth of experimental data which could be useful
in verifying some of the theoretical concepts.

The study of chaos in nuclear physics has  been mostly based --
so far -- on level statistics \cite{BFF}.
In practical terms this is rather restrictive
since a reliable statistical analysis requires very precise
energy resolution.
It also does not provide firm means for investigating the role of
collectivity and mechanisms of its coexistence with chaos \cite{GW}.
In this respect the study of temporal correlations between an initially
prepared non-stationary
state and a state to which it evolves seems to be much more appropriate.
Nuclear giant resonances are of central interest in this
connection because they are located in a region of high level density
which is expected to be dominated by chaotic dynamics.
To make the theoretical studies meaningful one needs a scheme
which incorporates the relevant elements, such as the possibility
of defining a physical collective state, a realistic modeling of the
background states whose complexity is consistent with
the Gaussian orthogonal
ensemble (GOE) of random matrices \cite{BFF} and,
finally, the realistic coupling between the two.

The recently developed model \cite{DNSW},
based on a diagonalization of the full nuclear Hamiltonian consisting of
a mean field part and a residual interaction
\begin{equation}
\hat H=\sum_i\epsilon_i a_i^{\dagger} a_i+{1\over 2}\sum_{ij,kl}v_{ij,kl}
a_i^{\dagger} a_j^{\dag} a_la_k,
\end{equation}
in the basis of 1p-1h and 2p-2h states
\begin{equation}
|1\rangle\equiv a_p^{\dagger} a_h|0\rangle;\quad |2\rangle\equiv
a_{p_1}^{\dagger} a_{p_2}^{\dagger}  a_{h_2}a_{h_1}|0\rangle
\end{equation}
fullfills these requirements and proves numerically manageable \cite{DNW}.
A prediagonalization of the two-body interaction $v$
in the 1p-1h and 2p-2h
subspaces defines $|\tilde 1\rangle=\sum_1C^{\tilde 1}_1|1\rangle$
and $|\tilde 2\rangle=\sum_2C^{\tilde 2}_2|2\rangle$, and the coupling is
mediated by the off-diagonal elements
$\langle \tilde 1|v|\tilde 2\rangle$ and their complex conjugate.
An initially excited state, in response to an external one-body field
$\hat F=\sum_{ij}F_{ij}a^{\dagger}_ia_j$ can be represented as
\begin{equation}
|F\rangle=\hat F |0 \rangle =\sum_{\tilde 1} F_{\tilde 1}
|{\tilde 1}\rangle.
\label{eq:F0}
\end{equation}
The amplitudes $F_{\tilde 1}=\langle {\tilde 1}|\hat F|0\rangle$
contain the entire information about the strength of a given spectral
line corresponding to the state $|{\tilde 1}\rangle$
and about the phase coherence among these states.
On the other hand, the level fluctuations
of the states $|\tilde 2\rangle$ can be used as a measure
of the degree of complexity of the background states \cite{DNSW}.
As soon as the coupling between the subspaces $|\tilde 1\rangle$ and
$|\tilde 2\rangle$ is taken into account, the state originally localized
in the 1p-1h subspace, as defined by eq.~(\ref{eq:F0}),
starts leaking into the 2p-2h space.
The degree of mixing depends not only on the
magnitude of $v$ but also on the nature of the energy fluctuations in
the 2p-2h space, which can significantly influence
the distribution of the coupling matrix elements \cite{DNW,MDVB}.
The most natural quantity for describing the leakage
is the survival probability, defined as
\begin{equation}
P(t)=|\langle F(0)|F(t) \rangle|^2.
\end{equation}
The time evolution of the state vector $|F(t)\rangle$ can be expanded
as
\begin{equation}
|F(t)\rangle =\sum_n a_n e^{-i E_n t/\hbar} |n\rangle,
\label{eq:tevol}
\end{equation}
where $E_n$ and $|n\rangle$ are the eigenenergies and eigenstates of
$\hat H$ in the space of 1p-1h and 2p-2h states. The
expansion coefficients are determined by the initial state as
$a_n=\langle n|F(0)\rangle$.
The physical significance of $P(t)$ can be identified from its relation
to the spectral autocorrelation function $G(E)$ \cite{LLJP,AL,LK}
\begin{equation}
P(t)=\int dE e^{-iEt/\hbar} G_F(E),
\label{eq:Four}
\end{equation}
where
\begin{equation}
G_F(E)=\int dE' S_F(E') S_F(E'+E)
\label{eq:Conv}
\end{equation}
and $S_F(E)$ is the transition strength distribution
$S_F(E)=\Sigma_n |F_n|^2 \delta (E-E_n)$.
Thus $P(t)$ can be obtained from experiments which measure
$S_F(E)$. Because of finite energy resolution the experiment
determines only an envelope of $S_F(E)$ and consequently
smoothes out the fluctuations in $P(t)$.
Average structures which are more interesting are preserved,
however, provided that the resolution is not too coarse.

The calculations presented below for quadrupole excitations in
$^{40}Ca$ are performed in the same basis as in ref.~\cite{DNW},
{\sl i.e.} including all 1p-1h and 2p-2h states up to 50 MeV
and using the same residual interaction. We then have 26 1p-1h and
3014 2p-2h states, which ensures a realistic description of the
transition strength distribution.
Since the present study concentrates on the phase-space exploration
and the role played by chaotic dynamics, we distinguish three cases
corresponding to different classes of the spectral fluctuations
in the 2p-2h subspace. As established in ref.~\cite{DNSW} one finds:
(a) with no residual interaction in the 2p-2h subspace there
are many degeneracies in $|\tilde 2\rangle$ ($=|2\rangle$)
and the nearest-neighbor spacing is strongly peaked near zero;
(b) inclusion of particle-particle and hole-hole matrix
elements in $\langle 2|v|2'\rangle$ removes all degeneracies
and leads to a Poissonian distribution of the nearest-neighbor spacings,
characteristic of generic integrable systems; and
(c) use of the full residual interaction yields GOE
fluctuations \cite{BFF}, characteristic of chaotic dynamics.

The initial state $|F(0)\rangle=\hat F|0\rangle$
(eq.~(\ref{eq:F0})) is already non-stationary in the 1p-1h subspace
and therefore oscillates within the limits set by this subspace.
The time evolution of the resulting survival probabilities
($|F\rangle$ is normalized to unity) are shown in
the upper panels of Figs.~1 and 2 for the isovector
and isoscalar quadrupole excitations, respectively.
It is interesting to note that the isoscalar excitation, being more
collective, overlaps - on average -  more frequently
with its initial state, as a comparison of the horizontal
lines in Figs.~1 and 2 indicates. Including the mixing with 2p-2h states
(making use of eq.~(\ref{eq:tevol}))
the results (Figs.~1 and 2) show that the isovector excitation
mixes much more efficiently
with the background 2p-2h states, both concerning the oscillatory
behavior and the asymptotic value of the survival probability $P(t)$.
The latter systematically decreases with increasing degree
of complexity in the background states (going from (a) to (c)).
Most interestingly, for the isovector excitation in the chaotic case (c),
$P(t)\cdot N$ where $N$ denotes the total number (3040) of states
in our space, reaches - on average - a value close to 3 (3.08).
It is known \cite{Hel}, that a state evolving from generic initial
conditions does not visit all the regions of the space with equal
probability but overlaps more frequently with its initial value.
This effect, characteristic of quantum ergodicity \cite{Hel},
is present even if the whole space is accessible.
In this extreme case it is just
the factor of 3 which prescribes the lowest limit on the average
asymptotic behavior of $P(t)$. Such an 'elastic enhancement'
finds empirical evidence in nuclear physics \cite{KW} and the factor
of 3 is considered as a quantum-mechanical
signature of chaos \cite{LQSL}. While the factor of 3 is consistent
with random-matrix-theory estimates \cite{UP}
the same asymptotic value may, in principle, be associated with
the regular dynamics \cite{WB} if the subspace is defined such that
there are no other conserved quantum numbers than those defining it.
An additional requirement for
chaoticity is the initial dephasing of $P(t)$ below its asymptotic
value \cite{LLJP,WB,AW,LK}. Such a dephasing, indeed, takes place for
the isovector case, as can be seen from Fig.~3 where
${\bar P(t)}=\int_0^t dt'P(t')/t$ for all three types of
the background spectra is displayed.
As a consequence of the high density of states the corresponding
'correlation hole' \cite{LLJP}
extends over a time interval four orders of magnitude longer than
the charactistic 'excitation time' of $\sim 10^{-22}$sec.
It should be emphasized that, contrary to all the previous
studies, no ensemble average over Hamiltonians or
initial conditions is present in our investigations.
The isoscalar excitation (right panel of Fig.~3) only
shows a trace of such a behavior and the asymptotic values of $P(t)$ are
systematically larger even though the initial state is
coupled to the same background.

The question what makes the isoscalar state so strongly
localized then arises.
Is it a manifestation of collectivity or, perhaps, is it
that the specific properties of the coupling matrix elements block
certain regions of the phase space and ergodization only occurs in
the unblocked regions?
A quantity which appears helpful in resolving this question
originates from the concept of entropy. The information entropy
of the state $|F\rangle$ can be defined, in a given basis $|k\rangle$,
as
\begin{equation}
K=-\sum_k p_k \ln p_k,
\label{eq:inf}
\end{equation}
where $p_k=|\langle k |F \rangle|^2$. It provides a quantitative measure
of the complexity of the state $|F\rangle$ and its localization length
in the basis $|k\rangle$ \cite{Izr}. The so-defined $K$ is,
in principle,
basis dependent, but the physically preferred basis is determined
by the mean field \cite{Zel1}.
The mean field, as the smoothest component of the nuclear Hamiltonian
\cite{Zel2}, provides a natural reference for quantifying the local
GOE-type fluctuations. In our case the mean field basis corresponds to
the unperturbed basis of states $|1\rangle$ and $|2\rangle$.
Calculating $K(t)$ along the 'trajectory' $|F(t)\rangle$
for the isovector and isoscalar states, we obtain asymptotically
values of  7.10 and 6.53, while the corresponding initial values are
2.56 and 2.40, respectively. This is to be compared to $K^{GOE}=7.29$
($K^{GOE}=\psi(N/2+1)-\psi(3/2)$ \cite{Izr},
where $\psi$ is the digamma function and $N$ is the number of basis
vectors). A comparison of these numbers indicates non-uniformities
in the $p_k$ distribution, especially for the isoscalar
excitations. Actually, even the GOE-type fluctuations result in a
gaussian distribution which is non-uniform (a uniform distribution
maximizes the entropy and corresponds to $p_k=1/N$ which, for
$N$=3040, gives $K \approx 8.02$).

In view of the above mentioned non-uniform phase-space
exploration, we find it instructive
to calculate the spectrum of $q$-moments for $\{p_k\}$ and to introduce
a generalized entropy \cite{Ren}
\begin{equation}
K_q= {1 \over{1-q}} \ln \sum_{k=1}^N p_k^q.
\label{eq:entr}
\end{equation}
{}From this definition it follows that $K_{q_1} \leq K_{q_2}$
if $q_2 < q_1$ (provided $\Sigma_k p_k =1$).
Equality holds for the uniform distribution. For $q \to 1$
eq.~(\ref{eq:entr}) yields the information entropy (eq.~(\ref{eq:inf})).
The most important property of $K_q$ is
that with increasing $q$ a higher weight is given to the largest
components in the set $\{p_k\}$. For $q \to 0$, on the other hand,
$K_q$ just counts the number of sites (here the basis vectors $|k\rangle$)
visited, irrespective of how frequently they are sampled.
For this reason eq.~(\ref{eq:entr}) also constitutes a basis for defining
the multifractal dimensions of non-uniform fractal sets \cite{GP}.

For selected $q$-values Fig.~4 compares the time evolution of
$K_q(t)$ for the isovector and isoscalar excitations when the background
states have GOE fluctuations (case (c)).
As one can see from the large-$q$ behavior of $K_q(t)$,
which are systematically smaller for the isoscalar excitation,
the large components of these remain much more localized (larger)
than those of the isovector excitation.
Since, by probability conservation,
the number of significant components is smaller
in the former case, the amplitude of oscillations is larger in the
corresponding $K_q(t)$.
On the other hand, the dynamics start to look similar in both cases as
$q$ decreases and, for $q \to 0$, $K_q$ approaches a  value of 8. This
signals that, on the level of small probabilities,
the whole space spanned by 3040 states is visited.
This aspect of the dynamics is consistent with the scaling
properties of the transition strength distribution for the isovector
and isoscalar states discussed in ref.~\cite{DNW}. On the level of small
components they both scale.
The similar calculation of $K_q(t)$ for the cases
(a) and (b) shows that the states evolve to configurations characterized
by significantly smaller entropies.

It would be very interesting to verify some of the above
predictions experimentally - especially the appearance
of the correlation hole in the average survival probability $\bar P(t)$.
This quantity is accessible through the convolution formula
(eq.~\ref{eq:Conv}) in conjunction with eq.~(\ref{eq:Four}).
In addition, for the purpose of addressing many specific questions
concerning quantum-mechanical phase space exploration, the full
set of generalized entropies appears to be very useful.

\vskip 1.5cm
\begin{center}
{\bf Acknowledgement}
\end{center}

This work was supported in part by the Polish KBN Grant
No. 2 P302 157 04 and by NSF Grant No. PHY-89-21025.
One of the authors (S. N.) would like to express his thanks to
the Alexander von Humboldt Foundation for a fellowship.

\newpage
\vspace{.25in}
\parindent=.0cm            %align refs

\newpage
\begin{center}
{\bf Figure Captions}
\end{center}

\begin{itemize}
\item[{\bf Fig.~1}] The time dependence of the
isovector quadrupole survival probability $P(t)$ in
$^{40}$Ca: (1p-1h) no coupling to the 2p-2h subspace;
(a) no residual interaction in 2p-2h subspace;
(b) including only particle-particle and
hole-hole matrixelements in the diagonalization of the 2p-2h subspace;
(c) diagonalization of the full residual interaction in the 2p-2h
subspace.
The solid horizontal lines indicate the time-averaged asymptotic values
for the corresponding $P(t)$.

\item[{\bf Fig.~2}] Same as Fig.~1 but for the isoscalar quadrupole
survival probability.

\item[{\bf Fig.~3}] The time-averaged survival probability
$\bar P(t)$ in units of $1/N$ (see text)
for the isovector (l.h.s.) and isoscalar (r.h.s.) quadrupole resonances:
(a) no residual interaction in 2p-2h subspace;
(b) including only particle-particle and
hole-hole matrix elements in the diagonalization of the 2p-2h subspace;
(c) diagonalization of the full residual interaction in the 2p-2h
subspace.

\item[{\bf Fig.~4}] The time evolution of the generalized entropies
defined by eq.~(\ref{eq:entr})
for the isovector (l.h.s) and isoscalar (r.h.s) giant resonances
corresponding to case (c) in Fig.~2. The horizontal marks on the left
hand side of each panel denote the asymptotic values of $K_q$ for
$q=4$ and $q=16$.

\end{itemize}


\begin{thebibliography}{99}
\bibitem{Hel} E.J. Heller, Phys. Rev. {\bf A35}, 1360(1987)
\bibitem{LLJP} L. Leviander, M. Lombardi, R. Jost and J.-P. Pique,
Phys. Rev. Lett. {\bf 56}, 2449(1986);\newline
J.-P. Pique, Y. Chen, R.W. Field and J.L. Kinsey,
Phys. Rev. Lett. {\bf 58}, 475(1987)
\bibitem{WB} J. Wilkie and P. Brumer, Phys. Rev. Lett. {\bf 67},
1185(1991)
\bibitem{LQSL} F. Leyvraz, J. Quezada, T.H. Seligman and M. Lombardi,
Phys. Rev. Lett. {\bf 67}, 2921(1991)
\bibitem{AW} Y. Alhassid and N. Whelan, Phys. Rev. Lett. {\bf 70},
572(1993)
\bibitem{BFF} T.A. Brody, J. Flores, J.B. French, P.A. Mello,
A. Pandey and S.S.M. Wong,
Rev. Mod. Phys. {\bf 53}, 385(1981);\newline
R.V.~Haq, A.~Pandey and O.~Bohigas,
Phys. Rev. Lett. {\bf 48}, 1086(1982);\newline
O. Bohigas, M.J. Giannoni and C. Schmit, Phys. Rev. Lett.
{\bf 52}, 1(1984)
\bibitem{GW} T. Guhr and H.A. Weidenm\"uller, Ann. Phys. {\bf 193},
472(1989)
\bibitem{DNSW} S. Dro\.zd\.z, S. Nishizaki, J. Speth and J. Wambach,
Phys. Rev. {\bf C49}, 867(1994)
\bibitem{DNW} S. Dro\.zd\.z, S. Nishizaki and J. Wambach,
Phys. Rev. Lett. {\bf 72}, 2839(1994)
\bibitem{MDVB} M. Matsuo, T. Dossing, E. Vigazzi and R.A. Broglia,
Phys. Rev. Lett. {\bf 70}, 2694(1993)
\bibitem{AL} Y. Alhassid and R.D. Levine, Phys. Rev. {\bf A46},
4650(1992)
\bibitem{LK} R.D. Levine and J.L. Kinsey, Proc. Natl. Acad. Sci.
{\bf 88}, 11133(1991)
\bibitem{KW} W. Kretschmer and M. Wangler, Phys. Rev. Lett. {\bf 41},
1224(1978)
\bibitem{UP} N. Ullach and C.E. Porter, Phys. Lett. {\bf 6}, 30(1963)
\bibitem{Izr} F.M. Izrailev, Phys. Rep. {\bf 196}, 299(1990)
\bibitem{Zel1} V.G. Zelevinsky, Nucl. Phys. {\bf A570}, 411c(1994)
\bibitem{Zel2} V.G. Zelevinsky, Nucl. Phys. {\bf A555}, 109(1993)
\bibitem{Ren} A. Renyi, {\it Probability Theory}
(North-Holland, Amsterdam, 1970)
\bibitem{GP} P. Grassberger and I. Procaccia, Physica {\bf 13D},
34(1984)
\end{thebibliography}
\end{document}